\documentclass[sigconf]{acmart}

\acmConference[ICSE 2022]{The 44th International Conference on Software Engineering}{May 21–29, 2022}{Pittsburgh, PA, USA}
\AtBeginDocument{%
  \providecommand\BibTeX{{%
    \normalfont B\kern-0.5em{\scshape i\kern-0.25em b}\kern-0.8em\TeX}}}

\setcopyright{acmcopyright}

\copyrightyear{2022}
\acmYear{2022}
\setcopyright{acmcopyright}\acmConference[ICSE '22]{44th International Conference
on Software Engineering}{May 21--29, 2022}{Pittsburgh, PA, USA}
\acmBooktitle{44th International Conference on Software Engineering (ICSE '22), May
21--29, 2022, Pittsburgh, PA, USA}
\acmPrice{15.00}
\acmDOI{10.1145/3510003.3510060}
\acmISBN{978-1-4503-9221-1/22/05}

\usepackage{multirow}
\usepackage{url} % add to cite url 
\usepackage{tablefootnote}
\usepackage{xspace}
\usepackage{multirow}
\usepackage{url} % add to cite url 
\usepackage{comment}
\usepackage{booktabs}
\newcommand{\myrevise}[3]{{\color{#2} #3}}
\newcommand\revise[1]{\myrevise{}{black}{#1}}
\newcommand{\Fig}{Figure\xspace}
\newcommand{\Tab}{Table\xspace}
\newcommand{\Sec}{Section\xspace}

\newcommand{\codenn}{CodeNN\xspace}
\newcommand{\deepcom}{Deepcom\xspace}

\newcommand{\astattgru}{Astattgru\xspace}
\newcommand{\rencos}{Rencos\xspace}

\newcommand{\NeuralCodeSum}{NCS\xspace}

\newcommand{\bleufc}{BLEU-FC\xspace} 
\newcommand{\sbleu}{Sentence BLEU\xspace}

\newcommand{\bleuncs}{BLEU-NCS\xspace} %\newcommand{\googlebleu}{Google-BLEU\xspace}
\newcommand{\bleurcs}{BLEU-RC\xspace} %\newcommand{\googlebleu}{rencos-BLEU\xspace}
\newcommand{\bleudc}{BLEU-DC\xspace}  %
\newcommand{\bleucn}{BLEU-CN\xspace} %\newcommand{\emsebleu}{EMSE-BLEU\xspace}
\newcommand{\bleu}{BLEU\xspace}
\newcommand{\cbleu}{Corpus BLEU\xspace}
\newcommand{\bleudcom}{BLEU-DM\xspace}
\newcommand{\avg}{Avg\xspace}

\newcommand{\tlcori}{TLC\xspace} % TLC\textsubscript{Original}
\newcommand{\tlcdedup}{TLC\textsubscript{Dedup}\xspace}

\newcommand{\csnfiltered}{CSN\xspace} % CSN\textsubscript{Filtered}
\newcommand{\csnpjtmed}{CSN\textsubscript{Project-Medium}\xspace}
\newcommand{\csnclsmed}{CSN\textsubscript{Class-Medium}\xspace}
\newcommand{\csnmtdmed}{CSN\textsubscript{Method-Medium}\xspace}
\newcommand{\csnmtdsmall}{CSN\textsubscript{Method-Small}\xspace}

\newcommand{\fcmfiltered}{FCM\xspace}  % FCM\textsubscript{Filtered}
\newcommand{\fcmpjtlarge}{FCM\textsubscript{Project-Large}\xspace}
\newcommand{\fcmmtdlarge}{FCM\textsubscript{Method-Large}\xspace}
\newcommand{\fcmmtdmed}{FCM\textsubscript{Method-Medium}\xspace}
\newcommand{\fcmmtdsmall}{FCM\textsubscript{Method-Small}\xspace}
\newcommand{\datasplit}{data splitting\xspace}

\newcommand{\processing}{pre-processing\xspace}

\newcommand{\nltk}{NLTK\xspace}
\newcommand{\nltkthreesix}{NLTK\textsubscript{3.6.x}\xspace}
\newcommand{\nltkthreefive}{NLTK\textsubscript{3.5.x}\xspace}

\newcommand{\nltkthreetwo}{NLTK\textsubscript{3.2.x}\xspace}

\newcommand{\nltkthreetwofour}{NLTK\textsubscript{3.2.4}\xspace}

\newcommand{\methodzero}{method\textsubscript{0}\xspace}

\newcommand{\methodtwo}{method\textsubscript{2}\xspace}
\newcommand{\methodthree}{method\textsubscript{3}\xspace}
\newcommand{\methodfour}{method\textsubscript{4}\xspace}

\newcommand{\rank}{ranking\xspace}

\newcommand{\preoooo}{$P_{0000}$\xspace} %0000
\newcommand{\preoool}{$P_{0001}$\xspace} %0001
\newcommand{\preoolo}{$P_{0010}$\xspace} %0010
\newcommand{\preooll}{$P_{0011}$\xspace} %0011
\newcommand{\preoloo}{$P_{0100}$\xspace} %0100
\newcommand{\preolol}{$P_{0101}$\xspace} %0101
\newcommand{\preollo}{$P_{0110}$\xspace} %0110
\newcommand{\preolll}{$P_{0111}$\xspace} %0111
\newcommand{\prelooo}{$P_{1000}$\xspace} %1000
\newcommand{\prelool}{$P_{1001}$\xspace} %1001
\newcommand{\prelolo}{$P_{1010}$\xspace} %1010
\newcommand{\preloll}{$P_{1011}$\xspace} %1011
\newcommand{\prelloo}{$P_{1100}$\xspace} %1100
\newcommand{\prellol}{$P_{1101}$\xspace} %1101
\newcommand{\prelllo}{$P_{1110}$\xspace} %1110
\newcommand{\prellll}{$P_{1111}$\xspace} %1111

\newcommand{\boxmargin}{0.5mm}

\usepackage{comment}
\usepackage{booktabs}
\usepackage{svg}
\usepackage{listings}
\usepackage{tcolorbox}
\usepackage{color}
\usepackage{xcolor}
\definecolor{dkgreen}{rgb}{0,0.6,0}
\definecolor{gray}{rgb}{0.5,0.5,0.5}
\definecolor{mauve}{rgb}{0.58,0,0.82}
\usepackage{colortbl}
\usepackage{lipsum}
\newcommand\equalauthorfootnote[1]{%
  \begingroup
  \renewcommand\thefootnote{}\footnote{\textsuperscript{\dag}#1}%
  \addtocounter{footnote}{-1}%
  \endgroup
}
\newcommand\corrauthorfootnote[1]{%
  \begingroup
  \renewcommand\thefootnote{}\footnote{\textsuperscript{\S}#1}%
  \addtocounter{footnote}{-1}%
  \endgroup
}
\begin{document}

\title{On the Evaluation of Neural Code Summarization}

\author{
Ensheng Shi\textsuperscript{a,\dag},
Yanlin Wang\textsuperscript{b,\S},
Lun Du\textsuperscript{b},
Junjie Chen\textsuperscript{c}\\
Shi Han\textsuperscript{b},
Hongyu Zhang\textsuperscript{d},
Dongmei Zhang\textsuperscript{b},
Hongbin Sun\textsuperscript{a,\S}
}
\affiliation{
\institution{\textsuperscript{a}Xi'an Jiaotong University \quad
\textsuperscript{b}Microsoft Research  }
\institution{\textsuperscript{c}Tianjin University \quad \textsuperscript{d}The University of Newcastle}
\country{}}
\email{ s1530129650@stu.xjtu.edu.cn, hsun@mail.xjtu.edu.cn}
\email{ {yanlwang, lun.du, shihan, dongmeiz}@microsoft.com } 
\email{{junjiechen@tju.edu.cn, hongyu.zhang@newcastle.edu.au}}

\renewcommand{\shortauthors}{Shi, et al.}

\tcbset{colback=gray!8,%gray background
        colframe=black,% black frame colour
        width=8.7cm,% Use 5cm total width,
        arc=2mm, auto outer arc,
        boxrule = 1.0pt,
        left = \boxmargin, right = \boxmargin, top = \boxmargin, bottom = \boxmargin,
        % left skip = 1mm
        leftright skip=0.5mm
}

\begin{abstract}
Source code summaries are important for program comprehension and maintenance. However, there are plenty of programs with missing, outdated, or mismatched summaries. Recently, deep learning techniques have been exploited to automatically generate summaries for given code snippets. To achieve a profound understanding of how far we are from solving this problem and provide suggestions to future research, in this paper, we conduct a systematic and in-depth analysis of 5 state-of-the-art neural code summarization models on 6 widely used BLEU variants, 4 pre-processing operations and their combinations, and 3 widely used datasets. The evaluation results show that some important factors have a great influence on the model evaluation, especially on the performance of models and the ranking among the models. However, these factors might be easily overlooked. Specifically, (1) the BLEU metric widely used in existing work of evaluating code summarization models has many variants. Ignoring the differences among these variants could greatly affect the validity of the claimed results. Besides, we discover and resolve an important and previously unknown bug in BLEU calculation in a commonly-used software package. Furthermore, we conduct human evaluations and find that the metric BLEU-DC is most correlated to human perception; (2) code pre-processing choices can have a large (from -18\% to +25\%)  impact on the summarization performance and should not be neglected. We also explore the aggregation of pre-processing combinations and boost the performance of models; (3) some important characteristics of datasets (corpus sizes, data splitting methods, and duplication ratios) have a significant impact on model evaluation. Based on the experimental results, we give actionable suggestions for evaluating code summarization and choosing the best method in different scenarios. We also build a shared code summarization toolbox to facilitate future research.
\end{abstract}



%
% Keywords. The author(s) should pick words that accurately describe
% the work being presented. Separate the keywords with commas.


\maketitle

% \blfootnote{Corresponding Author} %无标号，显示

\section{Introduction}

Source code summaries\equalauthorfootnote{Work performed during internship at Microsoft Research Asia.}\corrauthorfootnote{Corresponding authors.} are important for program comprehension and maintenance since developers can quickly understand a piece of code by reading its natural language description. However, documenting code with summaries remains a labor-intensive and time-consuming task. As a result, code summaries are often missing, mismatched, or outdated in many projects \cite{TilleyMO92,Briand03,ForwardL02}. Therefore, automatic generation of code summaries is desirable and many approaches have been proposed over the years~\cite{SridharaHMPV10, HaiducAM10ICSE, HaiducAMM10WCRE, EddyRKC13, RodegheroMMBD14}.

Recently, deep learning (DL) based models are exploited to generate better natural language summaries for code snippets~\cite{IyerKCZ16,HuLXLJ18,HuLXLLJ18,WanZYXY0Y18,hu2019deep,LeClairJM19,zhangretrieval20,AhmadCRC20}. These models usually adopt a neural machine translation framework to learn the alignment between code and summaries. Some studies also enhance DL-based models by incorporating information retrieval techniques~\cite{zhangretrieval20,WeiLLXJ20}. Generally, existing neural source code summarization models show promising results and claim their superiority over traditional approaches.  

However, we notice that in the current code summarization work, there are many important details that could be easily overlooked and important issues that have not received much attention. These details and issues are associated with evaluation metrics, evaluated datasets and experimental settings, and affect the evaluation and comparison of approaches. In this work, we would like to dive deep into the problem and answer: \textbf{\emph{how to evaluate and compare code summarization models more correctly and comprehensively?}} 

To answer the above question, we conduct systematic experiments of 5 representative code summarization approaches (including \codenn~\cite{IyerKCZ16}, \deepcom~\cite{HuLXLJ18}, \astattgru~\cite{LeClairJM19}, \rencos~\cite{zhangretrieval20} and \NeuralCodeSum~\cite{AhmadCRC20}) on 6 widely used \bleu variants, 4 extensively used code \processing operations (\Tab~\ref{tab:studies_diff_preprocessing}), and 3 commonly used datasets (including TL-CodeSum~\cite{HuLXLLJ18}, Funcom~\cite{LeClairJM19}, and CodeSearchNet~\cite{abs-1909-09436}). The 6 \bleu variants and 4 code \processing operations cover most of the studies on code summarization since 2010. Each dataset is used in at least 5 previous studies.

Our experiments can be divided into three major parts. 
First, we conduct an in-depth analysis of the \bleu metric, which is widely used in previous code summarization work~\cite{IyerKCZ16,HuLXLJ18,HuLXLLJ18,WanZYXY0Y18,hu2019deep,AlonBLY19,LeClairJM19,AhmadCRC20,zhangretrieval20,WeiL0FJ19,FengGTDFGS0LJZ20,LeClairHWM20,WeiLLXJ20} and perform human evaluations to find the \bleu variant that best correlates with human perception (\Sec~\ref{bleu_analysis}). 
Then, we study different code \processing operations in recent code summarization works and explore an ensemble learning based technique to boost the performance of code summarization models (\Sec~\ref{diff_process_method}). 
Finally, we conduct experiments on the three datasets from three perspectives: corpus sizes, \datasplit methods, and duplication ratios (\Sec~\ref{diff_dataset}). 
Through extensive experiments, we obtain the following major findings about the current neural code summarization evaluation.

The \textit{first major finding} is that there is a wide variety of BLEU metrics
used in prior work and they produce rather different results for the same generated summaries. 
Some previous studies~\cite{IyerKCZ16,AlonBLY19, LeClairJM19,zhangretrieval20,WeiLLXJ20,FengGTDFGS0LJZ20,HuLXLJ18,hu2019deep,WeiL0FJ19,lin2021improving} accurately describe the \bleu metric used and compare models under the same \bleu metric~\cite{IyerKCZ16,AlonBLY19, LeClairJM19,zhangretrieval20,WeiLLXJ20,FengGTDFGS0LJZ20,HuLXLJ18,hu2019deep,WeiL0FJ19,lin2021improving}. 
However, there are still many works~\cite{WanZYXY0Y18,HuLXLLJ18,fernandes2019structured,AhmadCRC20,WuZZ21} cite or describe inconsistent \bleu metrics, leading to confusion for subsequent research. 
What's worse, some software packages used in \cite{WeiL0FJ19, HuLXLJ18,hu2019deep} for calculating \bleu are \textit{buggy}: \textcircled{1} they may produce a BLEU score greater than 100\% (or even $>$ 700\%), which extremely exaggerates the performance of code summarization models, and \textcircled{2} the results are also different across different package versions. More importantly, \bleu scores between papers cannot be directly compared \cite{Post18}. 
However, some studies~\cite{AhmadCRC20,WuZZ21} copy the \bleu scores reported in other papers and directly compare with them under different   \bleu metrics. %In detail, 
For example, \cite{AhmadCRC20} copied the scores reported in ~\cite{WeiL0FJ19}, and \cite{WuZZ21} copied the scores reported in ~\cite{AhmadCRC20}. 
The \bleu implementations in their released code~\cite{WeiL0FJ19,AhmadCRC20} are different. Furthermore, the study~\cite{WuZZ21} does not release its source code.
Therefore, these studies may overestimate their model performance or may fail to achieve fair comparisons, even though they are evaluated on the same dataset with the same experimental setting.
Through human evaluation, we find that \textit{\bleudc} (\Sec~\ref{sec:bg_bleu}) correlates with human perception the most.
We further give some actionable suggestions on the usage of \bleu in \Sec~\ref{bleu_analysis}.

The \textit{second major finding} 
is that different \processing  combinations can affect the overall performance by a noticeable margin of -18\% to +25\%. 
The results of the exploration experiment show that a simple ensemble learning technique can boost the performance of code summarization models. 
We also give actionable suggestions on the choice and usage of code \processing operations in \Sec~\ref{diff_process_method}. 

The \textit{third major finding} is that code summarization approaches perform inconsistently on different datasets,  
i.e., one approach may perform better than other approaches on one dataset and poorly on another dataset. Furthermore, we experimentally find that three dataset attributes (corpus sizes, \datasplit methods, and duplication ratios) have important impact on the performance of code summarization models. We further give some suggestions about evaluation datasets in \Sec~\ref{diff_dataset}.

In summary, our findings indicate that in order to evaluate and compare code summarization models more correctly and comprehensively, we need to pay much attention to the implementation of \bleu metrics, the way of code \processing, and the usage of datasets. 
The major contributions of this work are as follows:
\begin{itemize}
\item We conduct an extensive evaluation of five representative neural code summarization models with different evaluation metrics, code \processing operations, and datasets.
\item We conduct human evaluation and find that \bleudc is most correlated to human perception for evaluating neural code summarization models among the six widely-used \bleu variants. 
\item We conclude that many existing code summarization models are not evaluated comprehensively and do not generalize well in new experimental settings. Therefore, more research is needed to further improve code summarization models.
\item Based on the evaluation results, we give actionable suggestions for evaluating code summarization models from multiple perspectives.
\item We build a shared code summarization toolbox\footnote{\url{https://github.com/DeepSoftwareAnalytics/CodeSumEvaluation}} containing 6 \bleu variants implementation, 4 code \processing operations and 16 of their combinations, 12 datasets, re-implementations of baseline approaches that do not have publicly available source code, and all experimental results described in this paper.
\end{itemize}

\section{Background}

\subsection{Code Summarization}
In the early stage of automatic source code summarization, template-based approaches~\cite{SridharaHMPV10, HaiducAM10ICSE, HaiducAMM10WCRE, EddyRKC13, RodegheroMMBD14} are widely used. However, a well-designed template requires expert domain knowledge. 
Therefore, information retrieval (IR) based approaches~\cite{HaiducAM10ICSE, HaiducAMM10WCRE,EddyRKC13, RodegheroMMBD14} are proposed. The basic idea is to retrieve terms from source code to generate term-based summaries or to retrieve similar source code and use its summary as the target summary. 
However, the retrieved summaries may not correctly describe the semantics and behavior of code snippets, leading to the mismatches between code and summaries.

Recently, Neural Machine Translation (NMT) based models are exploited to generate summaries for code snippets~\cite{IyerKCZ16,HuLXLJ18,HuLXLLJ18,WanZYXY0Y18,fernandes2019structured,hu2019deep,AlonBLY19,LeClairJM19,WeiL0FJ19,FengGTDFGS0LJZ20,AhmadCRC20,cai2020tag,bansal2021project,lin2021improving,xie2021exploiting,chen2018neural,LeClairHWM20,haque2020improved,ye2020leveraging,wang2021cocosum}.  
CodeNN~\cite{IyerKCZ16} is an early attempt that uses only code token sequences, followed by various approaches that utilize AST~\cite{HuLXLJ18,hu2019deep,AlonBLY19,LeClairJM19,LeClairHWM20,lin2021improving,Shi0D0HZS21}, 
API knowledge~\cite{HuLXLLJ18}, 
type information~\cite{cai2020tag}, 
global context~\cite{bansal2021project,haque2020improved,wang2021cocosum}, 
reinforcement learning~\cite{WanZYXY0Y18,wang2020reinforcement}, 
multi-task learning~\cite{xie2021exploiting}, dual learning~\cite{WeiL0FJ19,ye2020leveraging}, 
and pre-trained language models~\cite{FengGTDFGS0LJZ20}. In addition, hybrid approaches~\cite{zhangretrieval20,WeiLLXJ20} that combine the NMT-based and IR-based methods are proposed and shown to be promising.

\vspace{-3pt}
\subsection{BLEU}\label{sec:bg_bleu}
Bilingual Evaluation Understudy (\bleu)~\cite{PapineniRWZ02} is commonly used for evaluating the quality of the generated code summaries~\cite{IyerKCZ16,HuLXLJ18,HuLXLLJ18,WanZYXY0Y18,hu2019deep,AlonBLY19,LeClairJM19,AhmadCRC20,zhangretrieval20,WeiL0FJ19,FengGTDFGS0LJZ20,LeClairHWM20,WeiLLXJ20,haque2020improved,ye2020leveraging}.
In short, a \bleu score is a percentage number between 0 and 100 that measures the similarity between one sentence to a set of reference sentences using constituent n-grams precision scores. BLEU typically uses BLEU-1, BLEU-2, BLEU-3, and BLEU-4 (calculated by 1-gram, 2-gram, 3-gram, and 4-gram precisions) to measure the precision. A value of 0 means that the generated sentence has no overlap with the reference while a value of 100 means perfect overlap with the reference. Mathematically, the n-gram precision $p_n$ is defined as:
\begin{equation}\label{equ:bleu-cal2}
p_{n} = \frac{\sum_{C \in\{\text { Candidates }\}}{\sum_{n \text { -gram }\in \mathcal{C}} } \text { Count }_{\text {clip }}(n \text { -gram })}{\sum_{C^{\prime} \in\{\text { Candidates }\}} {\sum_{n \text { -gram }\in \mathcal{C}^{\prime}}}   \text { Count }\left(n \text { -gram }^{\prime}\right)}
\end{equation}

\noindent BLEU combines all n-gram precision scores using geometric mean:
\vspace{-5pt}
\begin{equation}
BLEU = BP \cdot \exp \sum\nolimits_{n=1}^{N} \omega_{n} \log p_{n}
\label{bleu_eq}
\end{equation}
$\omega_{n}$ is a uniform weight $1/N$ ($N=4$). The straightforward calculation will result in high scores for short sentences or sentences with repeated high-frequency n-grams. Therefore, Brevity Penalty (BP) is used to scale the score and each n-gram in the reference is limited to be used just once.

The original \bleu was designed for the corpus-level calculation~\cite{PapineniRWZ02}. 
For sentence-level BLEU, since the generated sentences and references are much shorter, $p_4$ is more likely to be zero when the sentence has no 4-gram or 4-gram match. Then the geometric mean will be zero even if $p_1$, $p_2$, and $p_3$ are large. In this case, the \bleu score correlates poorly with human judgment. Therefore, several smoothing methods are proposed~\cite{ChenC14} to mitigate this problem.

As \bleu can be calculated at different levels and with different smoothing methods, there are many \bleu variants used in prior work and they could generate
different results for the same generated summary. 
Here, we use the names of \bleu variants defined in ~\cite{GrosSDY20} and add another \bleu variant: \bleudcom, which is a \sbleu without smoothing~\cite{ChenC14} and is based on the implementation of \nltkthreetwofour. The meaning of these \bleu variants are: 
\begin{itemize}
    \item \bleucn: This is a Sentence \bleu metric used in \cite{IyerKCZ16,AlonBLY19,FengGTDFGS0LJZ20}. It applies a Laplace-like smoothing by adding 1 to both the numerator and denominator of $p_n$ for $n \geq 2$.
    
    \item \bleudcom : This is a Sentence BLEU metric used in \cite{HuLXLJ18}. It uses smoothing \methodzero based on \nltkthreetwofour.

    \item \bleudc: This is a Sentence BLEU metric based on \nltkthreetwofour smoothing \methodfour, used in \cite{hu2019deep,WeiL0FJ19}.
    
    \item \bleufc: This is an unsmoothed \cbleu metric based on \nltk 
    , used in \cite{LeClairJM19,LeClairHWM20,WeiLLXJ20}. 
    
    \item \bleuncs: This is a Sentence \bleu metric used in \cite{AhmadCRC20}. It applies a Laplace-like smoothing by adding 1 to both the numerator and denominator of all $p_n$.
    
    \item \bleurcs: This is an unsmoothed Sentence BLEU metric used in \cite{zhangretrieval20}. To avoid the divided-by-zero error, it adds a tiny number $10^{-15}$ in the numerator and a small number $10^{-9}$ in the denominator of $p_n$. 
\end{itemize}

There is an interpretation of \bleu scores by Google~\cite{googleBleuExp}, which is shown in \Tab~\ref{tab:interpretation_bleu}. 
We also show the original \bleu scores reported by existing approaches in \Tab~\ref{tab:sota_score}. These scores vary a lot. Specifically, 19.61 for \astattgru would be interpreted as ``hard to get the gist'' and 38.17 for \deepcom would be interpreted as ``understandable to good translations'' according to \Tab~\ref{tab:interpretation_bleu}. However, this interpretation is contrary to the results shown in ~\cite{LeClairJM19} where \astattgru is relatively better than \deepcom. To study this issue, we need to explore the difference and comparability of different metrics and experimental settings used in different works. 
\begin{table}[t]
\centering 
\small
\vspace{-5pt}
\caption{Interpretation of BLEU scores~\cite{googleBleuExp}.}
\begin{tabular}{ll}
\toprule
Score & Interpretation \\
\midrule
\textless 10	& {Almost useless}\\
10-19	& {Hard to get the gist}\\
20-29	& {The gist is clear, but has significant grammatical errors }\\
30-40	& {Understandable to good translations}\\
40-50	& {High quality translations}\\
50-60	& {Very high quality, adequate, and fluent translations}\\
\textgreater 60	& {Quality often better than human}\\
\bottomrule
\end{tabular}
\label{tab:interpretation_bleu}
\end{table}

\begin{table}[ht]
\centering
\renewcommand\tabcolsep{2.7pt}
\footnotesize
\caption{The best \bleu scores reported in their papers.}
\begin{tabular}{c ccccc}
\toprule
Model & \codenn~\cite{IyerKCZ16} 
& \deepcom~\cite{HuLXLJ18}  
& \astattgru~\cite{LeClairJM19}   
& \rencos~\cite{zhangretrieval20}  
& \NeuralCodeSum~\cite{AhmadCRC20}  \\
 \midrule
\bleu Score & 20.50 & 38.17& 19.61& 20.70& 44.14 \\
\bottomrule
\end{tabular}
\label{tab:sota_score}
\vspace{-10pt}
\end{table}

\section{Experimental Design}
\subsection{Datasets}

We conduct experiments on three widely used code summarization datasets: TL-CodeSum~\cite{HuLXLLJ18},  Funcom~\cite{LeClairJM19}, and CodeSearchNet~\cite{abs-1909-09436}. 

\underline{TL-CodeSum} has 87,136 method-summary pairs crawled from 9,732 Java projects created from 2015 to 2016 with at least 20 stars. The ratio of the training, validation and test sets is 8:1:1. 
Since all pairs are shuffled, there can be methods from the same project in the training, validation, and test sets.
In addition, there are exact code duplicates among the three partitions.

\underline{CodeSearchNet} is a well-formatted dataset containing 496,688 Java methods across the training, validation, and test sets. Duplicates are removed and the dataset is split into training, validation, and test sets in proportion with 8:1:1 by project (80\% of projects into training, 10\% into validation, and 10\% into testing) such that code from the same repository can only exist in one partition.

\underline{Funcom} is a collection of 2.1 million method-summary pairs from 28,945 projects. Auto-generated code and exact duplicates are removed. Then the dataset is split into three parts for training, validation, and testing with the ratio of 9:0.5:0.5 by project.

In \Sec~\ref{diff_dataset}, we find that the performance of the same model and the \rank among the models are different on different datasets. To study which characteristic (such as corpus size, deduplication, etc) of datasets affects the performance and how they affect the performance.
we modify some characteristics of the datasets and obtain 9 new variants. In total, we experiment on 12 datasets, as shown in \Tab~\ref{tab:three_dataset_all} the statistics. In this paper, we use TLC, FCM, and CSN to denote TL-CodeSum, Funcom, and CodeSearchNet, respectively. \tlcori is the original TL-CodeSum. \tlcdedup is a TL-CodeSum variant, which removes the duplicated samples from the testing set.
\csnfiltered and \fcmfiltered are CodeSearchNet and Funcom with source code that cannot be parsed by javalang\footnote{https://github.com/c2nes/javalang} filtered out. Javalang is used in many previous studies~\cite{zhangretrieval20,PanthaplackelNG20,PanthaplackelLG21,hu2019deep,lin2021improving,ZhangWZ0WL19} to parse source code.
\revise{The three magnitudes (small, medium and large) are defined by the training set size of three widely used datasets we investigated in this paper. Specifically, small: the training size of \tlcori, medium: the training size of \csnfiltered, large: the training size of \fcmfiltered.}
These datasets are mainly different from each other in corpus sizes, \datasplit ways, and duplication ratios. Their detailed descriptions can be found in \Sec~\ref{RQ_description}.

\begin{table*}[t]
\renewcommand\tabcolsep{4pt}
\centering 
\small
\caption{The statistics of the 12 datasets used. }
\begin{tabular}{llllllll}
\toprule
\multirow{2}{*}{Name} &\multicolumn{4}{c}{\#Method} &\multirow{2}{*}{\#Class} &\multirow{2}{*}{\#Project}  &\multirow{2}{*}{Description}\\
\cmidrule(r){2-5} 
& Training &Validation &Test &All \\
 \midrule
 \tlcori &69,708 &8,714 & 8,714 &87,136 &\quad--&9,732 &Original TL-CodeSum~\cite{HuLXLLJ18}\\
 \tlcdedup &69,708 &8,714  &6,449&84,871&\quad--&\quad-- &Deduplicated TL-CodeSum\\
  \midrule
 \csnfiltered &454,044 &15,299 &26,897&496,240 &136,495&25,596& Filtered CodeSearchNet~\cite{abs-1909-09436} \\ 
 \csnpjtmed   &454,044 &15,299 &26,897&496,240&136,495 &25,596  &\csnfiltered split by project\\
 \csnclsmed &448,368 &19,707 &28,165&496,240 &136,495&25,596     &\csnfiltered split by class \\
 \csnmtdmed &446,607 &19,855 &29,778 &496,240&136,495&25,596    &\csnfiltered split by method\\
 \csnmtdsmall &69,708 &19,855 &29,778&119,341 &\quad--&\quad--  &Subset of \csnmtdmed \\ 
  \midrule
  \fcmfiltered &1,908,694 & 104,948  &104,777&2,118,419&\quad--&28,790&Filtered Funcom~\cite{LeClairJM19}\\
  \fcmpjtlarge &1,908,694 & 104,948 &104,777&2,118,419&\quad--&28,790&Split \fcmfiltered{} by project\\
  \fcmmtdlarge  &1,908,694 & 104,948 &104,777&2,118,419&\quad--&28,790&Split \fcmfiltered{} by method\\
  \fcmmtdmed  &454,044  & 104,948 &104,777&663,769 &\quad--&\quad-- &Subset of \fcmmtdlarge\\
  \fcmmtdsmall  &69,708 & 104,948 &104,777&279,433 &\quad--&\quad-- &Subset of  \fcmmtdlarge\\
\bottomrule
\end{tabular}
\label{tab:three_dataset_all}
\end{table*}

\subsection{Evaluated Approaches}

We choose the five approaches with the consideration of representativeness and diversity.

\underline{\codenn{}}~\cite{IyerKCZ16} is the first neural approach that
    learns to generate summaries of code snippets. It is a classical 
    encoder-decoder framework in NMT that encodes
    code to context vectors and then generates
    summaries in the decoder with the attention mechanism.

\underline{\deepcom{}}~\cite{HuLXLJ18} is an SBT-based (Structure-based Traversal) model, which can capture the syntactic and structural information from AST. It is an attentional LSTM-based encoder-decoder neural network that encodes the SBT sequence and generates summaries.
    
\underline{\astattgru{}}~\cite{LeClairJM19} is a
    multi-encoder neural model that encodes both code and AST to learn lexical and syntactic information of Java methods. It uses two GRUs to encode code and SBT sequences, respectively.

\underline{\NeuralCodeSum{}}~\cite{AhmadCRC20} is the first attempt to replace the previous RNN units with the more advanced Transformer model, and it incorporates the copying mechanism~\cite{SeeLM17} in the Transformer to allow both generating words from vocabulary and copying from the input source code.

\underline{\rencos{}}~\cite{zhangretrieval20} is a representative model that combines information retrieval techniques with the generation model in the code summarization task. Specifically, it enhances the neural model with the most similar code snippets retrieved from the training set. 

\subsection{Experimental Settings}
We use the default hyper-parameter settings provided by
each method and adjust the embedding size, hidden size, learning rate, and max epoch empirically to ensure that each model performs well on each dataset. We adopt max epoch 200 for \tlcori{} and \tlcdedup(others are 40) and early stopping with patience 20 to enable the convergence and generalization. In addition, we run each experiment 3 times and display the mean and standard deviation in the form of $mean\pm std$. All experiments are conducted on a machine with 252 GB main memory and 4 Tesla V100 32GB GPUs. 

We use the provided implementations by each approach: 
\codenn\footnote{\url{https://github.com/sriniiyer/codenn}}, 
\astattgru\footnote{\url{https://bit.ly/2MLSxFg}}, 
\NeuralCodeSum\footnote{\url{https://github.com/wasiahmad/NeuralCodeSum}} and 
\rencos\footnote{\url{https://github.com/zhangj111/rencos}}. 
For \deepcom, we re-implement the method\footnote{The code for our re-implementation is included in our toolbox.} according to the paper description since it is not publicly available. We have checked the correctness by reproducing the scores in the original paper~\cite{HuLXLJ18} and double confirmed with the authors of \deepcom.

\subsection{Research Questions}\label{RQ_description}

We investigate three research questions from three aspects: metrics, \processing operations, and datasets.

\noindent\textbf{RQ1: How do different \bleu variants affect the evaluation of code summarization?}

There are several metrics commonly used for various NLP tasks such as machine translation, text summarization, and captioning. These metrics include BLEU~\cite{PapineniRWZ02}, Meteor~\cite{BanerjeeL05},
Rouge-L~\cite{lin-2004-rouge},  Cider~\cite{VedantamZP15}, etc. In RQ1, we only present BLEU as it is the most commonly used metric in the code summarization task. To study "how do different \bleu variants affect the evaluation of code summarization?" and find "which variant should we use in practice?", we conduct some extensive experiments and the human evaluation. We first train and test the 5 approaches on \tlcori and \tlcdedup, and measure their generated summaries using different \bleu variants. Then we will introduce the differences of the \bleu variants in detail, and summarize the reasons for the differences from three aspects: different calculation levels (sentence-level v.s. corpus-level), different smoothing methods used, and many problematic software implementations.
Finally, we analyze the impact of each aspect, conduct human evaluation, and provide actionable guidelines on the use of \bleu, such as how to choose a smoothing method, 
and how to report \bleu scores more clearly and comprehensively.

\textit{Human evaluation.} To find which \bleu correlates with the human perception the most, we conduct a human evaluation. First, we randomly sample 300 (100 per dataset) generated summaries paired with original summaries. Then, we invite 5 annotators with excellent English ability and more than 2 years of software development experience. 
Each annotator is asked to assign scores from 0 to 4 to measure the semantic similarity between reference and generated summaries. The detailed meaning of these scores is given in \Tab 1 of our online Appendix\footnote{https://github.com/DeepSoftwareAnalytics/CodeSumEvaluation/tree/master/Appendix}.
To verify the agreement among the annotators, we calculate the 
Krippendorff’s alpha \cite{hayes2007answering} and Kendall rank correlation coefficient (Kendall’s Tau) \cite{kendall1945treatment} values. 
The value of Krippendorff’s alpha is 0.93, and the values of pairwise Kendall’s Tau range from 0.87 to 0.99, which indicates that there is a high degree of agreement between the 5 annotators and the scores are reliable. Then, we average scores of  5 annotators as the human score for each generated summary. Finally, following Wei at al.~\cite{abs-2107-05373},  we use Kendall’s rank correlation coefficient $\tau$~\cite{kendall1945treatment} and Spearman correlation coefficient $\rho$ \cite{ziegel2001standard} to measure the correlation between the human evaluation and each \bleu variant. 

\emph{Human score for each corpus.}
To study the correlation between \bleu variants and human evaluation at the corpus-level, we should obtain the human score of a corpus.
Following \cite{MaWBG19}, we average the human scores over all generated summaries as the final human score for a corpus. 
We use both arithmetic and geometric average in this paper. 

\emph{Number of summaries in each corpus.} 
To ensure the generalization and reliability of the conclusion, we randomly sample $x$ summaries from 300 scored samples as a corpus, where $x \in \{1, 20, 40, 60, 80, 100\}$, and we repeat this sampling process 5000 times. 

\noindent\textbf{RQ2: How do different \processing operations affect the performance of code summarization?}

\begin{table}[t]
\centering
\small
\caption{Code \processing operations used in previous code summarization work.}
\label{tab:studies_diff_preprocessing}
\begin{tabular}{c p{2.3cm} p{3.8cm}}
\toprule
Operation & Studies & Meaning\\  \midrule
$R$ & \cite{HuLXLJ18,hu2019deep,HuLXLLJ18,lin2021improving,WeiL0FJ19,WuZZ21} 
    & Replace string/number with generic symbols \texttt{<STRING>}/\texttt{<NUM>} \\\midrule
$S$ & \cite{AhmadCRC20,AllamanisPS16,AlonBLY19,bansal2021project,fernandes2019structured,HaiducAM10ICSE,haque2020improved,hu2019deep,HuLXLLJ18,LeClairHWM20,LeClairJM19,SridharaHMPV10,wang2020reinforcement,WeiL0FJ19,WeiLLXJ20,WuZZ21,zhangretrieval20} 
    & Split tokens using camelCase and snake\_case\\  \midrule
$F$ & \cite{bansal2021project,haque2020improved,LeClairHWM20,LeClairJM19,WeiLLXJ20} 
    & Filter the punctuations in code\\  \midrule
$L$& \cite{AllamanisPS16,bansal2021project,haque2020improved,hu2019deep,HuLXLLJ18,LeClairHWM20,LeClairJM19,lin2021improving,wang2020reinforcement,WeiL0FJ19,WeiLLXJ20,WuZZ21}
    & Lowercase all tokens \\  \midrule
Others & \cite{chen2018neural,IyerKCZ16,FengGTDFGS0LJZ20,MorenoASMPV13,RodegheroMMBD14,WanZYXY0Y18,xie2021exploiting} & No \processing, BPE, etc \\  
\bottomrule
\end{tabular}
\end{table}

There are various code \processing operations used in related work, such as token splitting, lowercase. We study recent papers on code summarization since 2010  according to the \processing operations they have used and summarize the result in \Tab~\ref{tab:studies_diff_preprocessing}. 
We select four operations $R,S,F,L$ that are most widely used to investigate whether different \processing operations would affect performance and find the dominated \processing choice.

We define a bit-wise notation $P_{RSFL}$ to denote different \processing combinations. For example, \prelolo means $R=True$, $S=False$, $F=True$, and $L=False$, which stands for performing $R$, $F$, and preventing $S$, $L$.
Then, we evaluate different \processing combinations on \tlcdedup dataset in \Sec~\ref{diff_process_method}.

\noindent\textbf{RQ3: How do different characteristics of datasets affect the performance?}

Many datasets have been used in source code summarization. 
We first evaluate the performance of different methods on three widely used datasets, which are different in three attributes: corpus sizes, \datasplit methods, and duplication ratios. Then, we study the impact of the three attributes with the extended datasets shown in \Tab~\ref{tab:three_dataset_all}.
The three attributes we consider are as follows:
% \begin{itemize}[topsep=0pt,itemsep=0pt,partopsep=0pt,parsep=0pt,leftmargin=8pt]

\underline{Data splitting methods}: there are three \datasplit ways we investigate: %used in code summarization:
\textcircled{1} by method: randomly split the dataset after shuffling the all samples~\cite{HuLXLLJ18},  \textcircled{2} by class: randomly divide the classes into the three partitions such that code from the same class can only exist in one partition, 
and \textcircled{3} by project: randomly divide the projects into the three partitions such that code from the same project can only exist in one partition~\cite{abs-1909-09436,LeClairJM19}.

\underline{Corpus sizes}: there are three magnitudes of training set size we investigate: 
\textcircled{1} small: the training size of \tlcori, 
\textcircled{2} medium: the training size of \csnfiltered, 
and \textcircled{3} large: the training size of \fcmfiltered.

\underline{Duplication ratios}: 
Code duplication is common in software development practice.  
This is often because developers copy and paste code snippets and source files from other projects~\cite{lopes2017dejavu}. According to a large-scale study~\cite{mockus2007large}, more than 50\% of files were reused in more than one open-source project. Normally, for evaluating neural network models, the training set should not contain samples in the test set. Thus, ignoring code duplication may result in model performance and generalization ability not being comprehensively evaluated according to the actual practice.
Among the three datasets we experimented on, Funcom and CodeSearchNet contain no duplicates because they have been deduplicated, but we find the existence of 20\% exact code duplication in TL-CodeSum. 
Therefore, we conduct experiments on TL-CodeSum with different duplication ratios to study this effect.

\section{Experimental Results}

\subsection{Analysis of Different Evaluation Metrics. (\textbf{RQ1})}\label{bleu_analysis}

\begin{table*}[ht]
\centering  
\small
\caption{Different metric scores in \tlcori and \tlcdedup. Underlined \underline{scores} refer to the metric used in the corresponding papers.} 
    \begin{tabular}{c cccccc cccccc}
    \toprule
    
        \multirow{3}{*}{Model} & \multicolumn{6}{c}{\tlcori}& \multicolumn{6}{c}{\tlcdedup}\\

    \cmidrule(lr){2-7}\cmidrule(r){8-13}
 &\footnotesize{\bleudcom}&\footnotesize{\bleufc} & \footnotesize{\bleudc} &\footnotesize{\bleucn}&\footnotesize{\bleuncs}&\footnotesize{\bleurcs}&\footnotesize{\bleudcom}&\footnotesize{\bleufc} & \footnotesize{\bleudc}&\footnotesize{\bleucn}&\footnotesize{\bleuncs}&\footnotesize{\bleurcs}\\ 
           &$s,m_0$  & $c,m_0$ & $s,m_4$ &$s,m_2$   &$s,m_l$&$s,m_0$ &$s,m_0$  & $c,m_0$ & $s,m_4$ &$s,m_2$   &$s,m_l$ &$s,m_0$\\
          \midrule
        \codenn &51.98 &26.04  &36.50 & \underline{33.07} &33.78  &26.32  &40.95  &8.90  & 20.51   &\underline{15.64} &16.60 &7.24 \\ 
        
        \deepcom &\underline{40.18} &12.14 &24.46 & 21.18 & 22.26 &13.74 &\underline{34.81} &4.03  & 15.87  &11.26 &12.68 &3.51  \\ 
        
        \astattgru &50.87 &\underline{27.11} &35.77& 31.98 &32.64 &25.87 &38.41 & \underline{7.50}  &18.51   & 13.35  &14.24 & 5.53  \\ 
        
        \rencos  &58.64&41.01&47.78&46.75&47.17&\underline{40.39}  &45.69 &22.98  &31.22  &29.81  &30.37 &\underline{21.39} \\
        
        \NeuralCodeSum &57.08 &36.89&45.97&45.19 & \underline{45.51}&38.37 & 43.91  &18.37 &29.07  & 27.99   &\underline{28.42} &18.94  \\
    \bottomrule
    \multicolumn{13}{l}{\footnotesize $s$ and $c$ represent sentence \bleu and corpus \bleu,  respectively. $m_x$ represents different smoothing methods, }\\
     \multicolumn{13}{l}{\footnotesize $m_0$ is without smoothing method, and $m_l$ means using add-one Laplace smoothing which is similar to $m_2$.}\\
    \end{tabular}

\label{tab:diff_metric}
\end{table*}

We experiment on the five approaches and measure their generated summaries using different \bleu variants. The results are shown in \Tab~\ref{tab:diff_metric}. We can find that:
\begin{itemize}
\item The scores of different \bleu variants are different for the same summary. For example, the \bleu scores of \deepcom on \tlcori{} vary from 12.14 to 40.18. \astattgru is better than \deepcom in all \bleu variants.

\item The \rank of models is not consistent using different \bleu variants. For example, the score of \astattgru is higher than that of \codenn in terms of \bleufc but lower than that of \codenn in other \bleu variants on \tlcori. 
\item Under the \bleufc measure, many existing models (except \rencos) have scored lower than 20 on \tlcdedup dataset. According to the interpretations in \Tab~\ref{tab:interpretation_bleu}, this means that under this experimental setting, the generated summaries are not gist-clear and understandable.

\end{itemize}

Next, we elaborate on the differences among the \bleu variants. The mathematical equation of \bleu is shown in Equation \eqref{bleu_eq},
which combines all n-gram precision scores using the geometric mean. The BP (Brevity Penalty) is used to scale the score because the short sentence such as single word
outputs could potentially have high precision.

\bleu~\cite{PapineniRWZ02} is firstly designed for measuring the generated corpus; as such, it requires no smoothing, as some sentences would have at least one n-gram match. For sentence-level \bleu, $p_4$ will be zero when the example has not a 4-gram, and thus the geometric mean will be zero even if $p_n(n < 4)$ is large. For sentence-level measurement,  it usually correlates poorly with human judgment. Therefore, several smoothing methods have been proposed in ~\cite{ChenC14}. 
\nltk\footnote{https://github.com/nltk/nltk} (the Natural Language Toolkit), which is a popular toolkit with 9.7K stars, implements the corpus-level and sentence-level measures with different smoothing methods and are widely used in evaluating generated summaries~\cite{HuLXLJ18,hu2019deep,WeiL0FJ19,HuLXLLJ18,LeClairJM19,LeClairHWM20,StapletonGLEWL020,WeiLLXJ20}. However, there are problematic implementations in different \nltk versions, leading to some \bleu variants unusable. 
We further explain these differences in detail.

\subsubsection{Sentence v.s. corpus BLEU}
The \bleu score calculated at the sentence level and corpus level is different, 
which is mainly caused by the different calculation strategies for merging all sentences. 
The corpus-level BLEU treats all sentences as a whole, where the numerator of $p_n$ is the sum of the numerators of all sentences' $p_n$, and the denominator of $p_n$ is the sum of the denominators of all sentences' $p_n$. Then the final \bleu score is calculated by the geometric mean of $p_n (n=1,2,3,4)$. Different from corpus-level \bleu, sentence-level \bleu is calculated by separately calculating the \bleu scores for all sentences, and then the arithmetic average of them is used as sentence-level \bleu. 
In other words, sentence-level \bleu aggregates the contributions of each sentence equally, while for corpus-level, the contribution of each sentence is positively correlated with the length of the sentence. 
Because of the different calculation methods, the scores of the two are not comparable. We thus suggest explicitly report at which level the \bleu is being used.

\subsubsection{Smoothing methods}
\begin{figure}[ht]
    \centering
    \begin{minipage}[ht]{1\linewidth}
        \centering
        \includegraphics[width=0.98\linewidth]{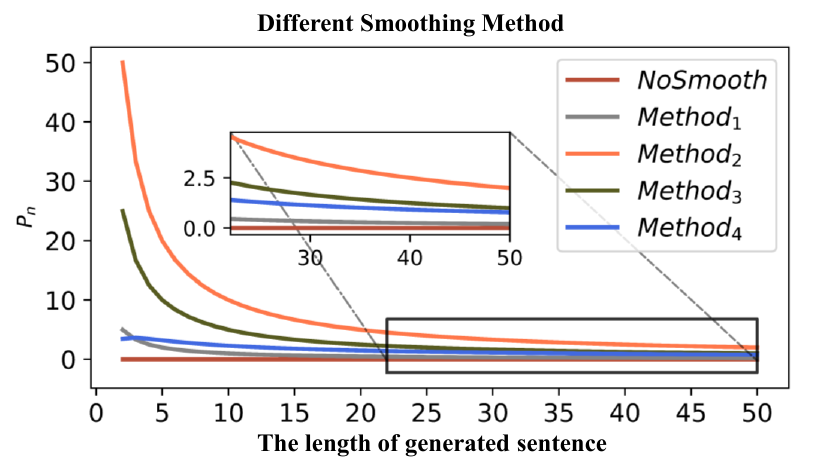}
    \end{minipage}
    
    \begin{minipage}[ht]{1.0\linewidth}
        \centering
          \includegraphics[width=0.98\linewidth]{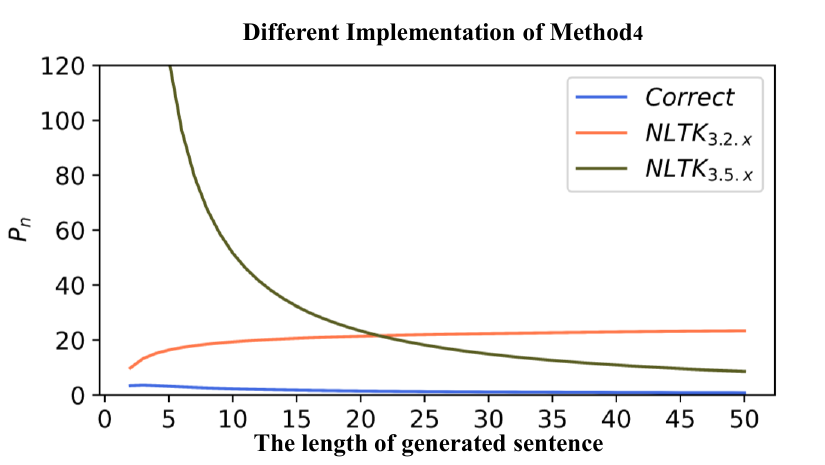}
    \end{minipage}%
\caption{Comparison of different smoothing methods.}
\label{fig:diff_smoothing_method}
\end{figure}

Smoothing methods are applied when deciding how to deal with cases if the number of matched n-grams is 0. Since \bleu combines all n-gram precision scores ($p_n$) using the geometric mean, \bleu will be zero as long as any n-gram precision is zero. One may add a small number to $p_n$, however, it will result in the geometric mean being near zero. Thus, many smoothing methods are proposed. Chen et al.~\cite{ChenC14} summarized 7 smoothing methods.
Smoothing methods 1-4 replace 0 with a small positive value, which can be a constant or a function of the generated sentence length. 
Smoothing methods 5-7 average the $n-1$, $n$, and $n+1$–gram matched counts in different ways to obtain the n-gram matched count. 
We plot the curve of $p_n$ under different smoothing methods applied to sentences of varying lengths in \Fig~\ref{fig:diff_smoothing_method} (upper). We can find that the values of $p_n$ calculated by different smoothing methods can vary a lot, especially for short sentences, which are often seen in code summaries. 

\subsubsection{Bugs in software packages} 
We measure the same summaries generated by \codenn in three \bleu variants (\bleudcom, \bleufc, and \bleudc), which are all based  on the \nltk implementation (but with different versions). From \Tab~\ref{tab:diff_nltk_v_dup}, we can observe that scores of  \bleudcom and \bleudc are very different under different \nltk versions (from ${3.2.x}$ to ${3.5.x}$).  
This is because the buggy implementations for \methodzero and \methodfour in different versions, which can cause up to 97\% performance difference for the same metric.

\noindent \textbf{Smoothing \methodzero bug.} 
\methodzero (means no smoothing method) of \nltkthreetwo only combines the \textbf{non-zero} precision values of all n-grams using the geometric mean. For example, \bleu is the geometric mean of $p_1$, $p_2$, and $p_3$ when $p_4=0$ and $p_n \neq 0 (n=1,2,3)$. 

\begin{figure}[t]
    \centering
    \includegraphics[width=0.99\linewidth]{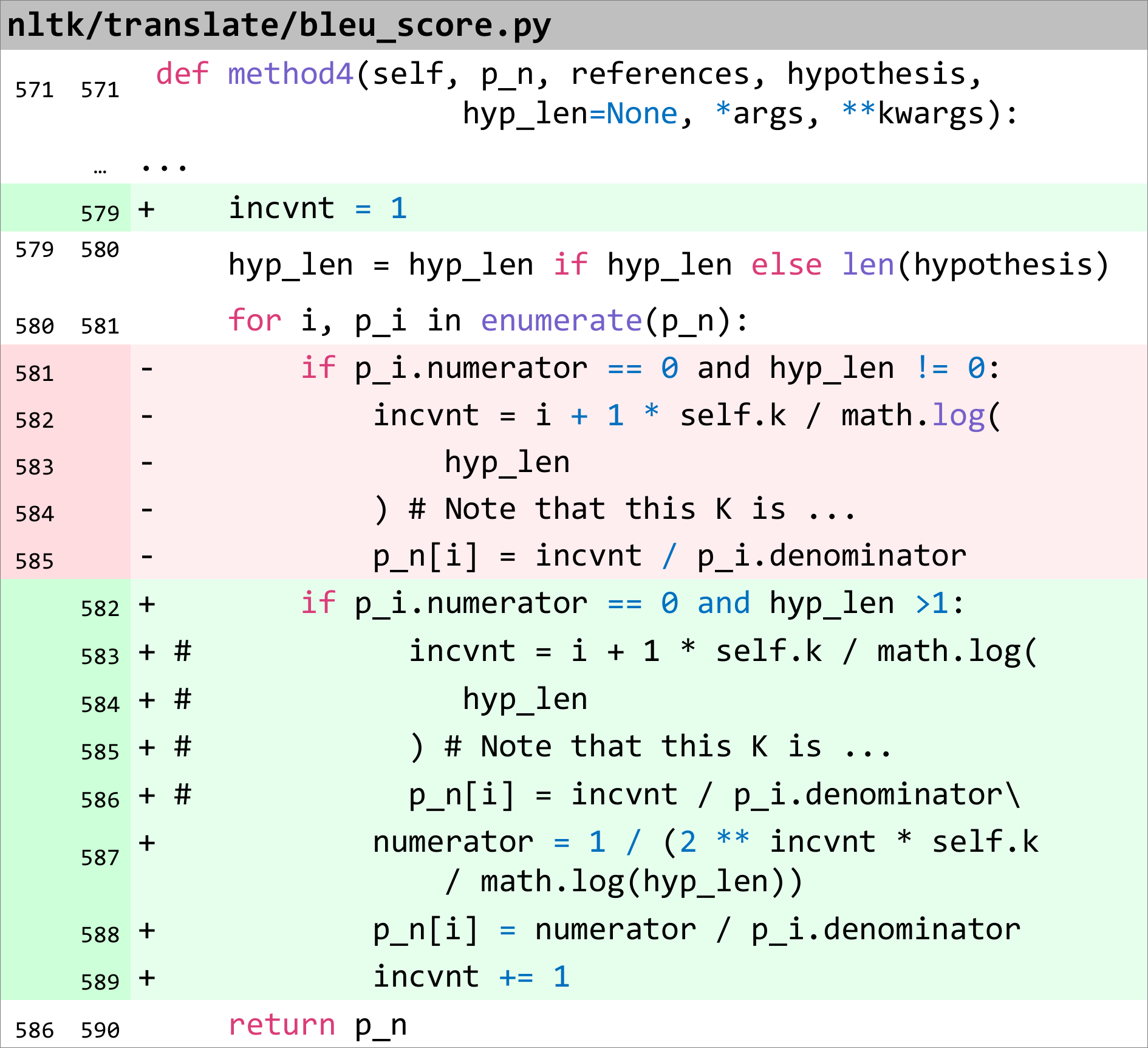}
    \caption{Issue 2676\textsuperscript{\ref{nltk_issue}} about smoothing method\textsubscript{4} in \nltk, which is reported and fixed by us.}
    \label{fig:nltk_method4_revise}
\end{figure}

\noindent \textbf{Smoothing \methodfour bugs.} 
\methodfour is implemented problematically in different \nltk versions. 
We plot the curve of $p_n$ of different smoothing \methodfour implementations in \nltk in \Fig~\ref{fig:diff_smoothing_method} bottom, where the correct version is \nltkthreesix.
In \nltk versions 3.2.2 to 3.4.x, $p_n = \frac{1} {n-1 + C/ln(l_h)}$, where $C=5$,  which always inflates the score in different length (\Fig~\ref{fig:diff_smoothing_method}).
The correct \methodfour proposed in \cite{ChenC14} is 
$p_n = 1 / {(invcnt * \frac{C}{\ln (l_h)} * l_h)}$
, where $C=5$ and $invcnt=\frac{1}{2^k}$ is a geometric sequence starting from 1/2 to n-grams with 0 matches.
In \nltkthreefive, $p_n = \frac{n-1 + 5/ln (l_h)}{l_h}$ where $l_h$ is the length of the generated sentence, thus $p_n$ can be assigned with a percentage number that is much greater than 100\% (even $>$ 700\%) when $l_h$ $<$ 5 in n-gram.
We have reported this issue\footnote{\url{https://github.com/nltk/nltk/issues/2676}\label{nltk_issue}} and filed a pull request\footnote{\url{https://github.com/nltk/nltk/pull/2681}} to \nltk GitHub repository, which has been accepted and merged into the official \nltk library and released in \nltkthreesix (the revision is shown in \Fig~\ref{fig:nltk_method4_revise}). Therefore, \nltkthreesix should be used when using smoothing \methodfour.

From the above experiments, we can conclude that \bleu variants used in prior work on code summarization are different from each other and the differences can carry some risks such as the validity
of their claimed results.
Thus, it is unfair and risky to compare different models without using the same \bleu implementation. For instance, it is unacceptable that researchers ignore the differences among the \bleu variants and directly compare their results with the  \bleu scores reported in other papers. 
We use the correct implementation to calculate \bleu scores in the following experiments. 

\begin{table}[t]
\centering \small
\caption{\bleu scores in different \nltk versions.} 
\begin{tabular}{c ccc|c}
  \toprule
  \multirow{2}{*}{Metric} & \multicolumn{4}{c}{\nltk version}\\
  \cmidrule(r){2-5}
        &${3.2.x}$\tablefootnote{Except for versions 3.2 and 3.2.1, as these versions are buggy with the \texttt{ZeroDivisionError} exception. Please refer to  \url{https://github.com/nltk/nltk/issues/1458}\label{nltk32_bug} for more details.} 
        &${3.3.x}$/$3.4.x$ 
        &${3.5.x}$  
        &${3.6.x}$\tablefootnote{\nltkthreesix are the versions with the BLEU calculation bug fixed by us.}  \\
    \midrule
     \bleudcom$(s,m_0)$ &51.98  &26.32 &26.32 &26.32 \\
       \bleufc $(c,m_0)$  &26.04  &26.04  &26.04  &26.04 \\
     \bleudc $(s,m_4)$  &36.50  &36.50 &42.39 &28.35\\  
    \bottomrule
    \end{tabular}
\label{tab:diff_nltk_v_dup}
\end{table}

\subsubsection{Human evaluation} 

\begin{table}[t]
\renewcommand\tabcolsep{1.5pt}
\caption{The values of correlation coefficients. \revise{$\rho$ is Spearman correlation coefficient and $\tau$ is Kendall rank correlation coefficient}. Here we use arithmetic average to aggregate summary-level human score as the corpus-level score. \revise{All results are statistically significant ($p \ll 0.05$)}.} 
\centering 
\footnotesize
\begin{tabular}{ccccccccccccc}
\toprule
\multirow{2}{*}{Metric}& \multicolumn{2}{c}{1} & \multicolumn{2}{c}{20} & \multicolumn{2}{c}{40 }& \multicolumn{2}{c}{60 }& \multicolumn{2}{c}{80 } & \multicolumn{2}{c}{100 }  \\ 
\cmidrule(r){2-3} \cmidrule(r){4-5} \cmidrule(r){6-7} \cmidrule(r){8-9} \cmidrule(r){10-11} \cmidrule(r){12-13}
& $\tau$  &$\rho$  & $\tau$  &$\rho$   & $\tau$  &$\rho$  & $\tau$  &$\rho$  & $\tau$  &$\rho$ & $\tau$  &$\rho$ \\
\midrule
\bleudcom $s,m_0$ &0.32&0.68&0.61 &0.80  &0.62 &0.81 &0.62 &0.81 &0.62 &0.82 &0.61 &0.8\\
\bleufc $c,m_0$  &0.32&0.68  &0.41 &0.58 &0.39 &0.56  &0.38 &0.55 &0.38 &0.55 &0.37 &0.54\\
\bleudc $s,m_4$   &\textbf{0.54} &\textbf{0.75} &\textbf{0.65} &\textbf{0.84} &\textbf{0.66} &\textbf{0.85} &\textbf{0.66} &\textbf{0.85}  &\textbf{0.66} &\textbf{0.85} &\textbf{0.65} &\textbf{0.84}\\
\bleucn $s,m_2$ &0.47&0.66  &0.60 &0.79&0.61 &0.81&0.62 &0.81 &0.62 &0.81 &0.61 &0.81\\
\bleuncs $s,m_l$ &0.37&0.53 &0.57 &0.76 &0.58 &0.78 &0.59 &0.78  &0.59 &0.79 &0.58 &0.78\\
\bleurcs $s,m_0$ &0.32 &0.68  &0.61 &0.80  &0.62 &0.81 &0.62 &0.81 &0.62 &0.82 &0.61 &0.8\\
\bottomrule
\end{tabular}
\label{tab:human_eval_avg}
\end{table}

To answer the question ``which \bleu correlates with human perception the most'', we conduct the human evaluation. \Tab~\ref{tab:human_eval_avg} shows the values of correlation coefficient under different corpus sizes when using arithmetic average\footnote{We also conduct another experiment that uses a geometric average to aggregate summary-level human scores as the corpus-level score. As the conclusion is consistent with the arithmetic average experiment, we put the results in the online Appendix Table 2 to save space.} to aggregate summary-level human scores as the corpus-level score.  \Tab\ref{tab:human_eval_avg} shows that, in terms of correlation coefficient $\tau$, 
\textcircled{1} when the corpus size is 1 (one-sentence level), \bleu metrics without smoothing method (\bleudcom, \bleufc, and \bleurcs) correlate poorly with human perception, and smoothing methods improve the correlation over no smoothing. Both findings are consistent with previous studies~\cite{RoyFA21,ChenC14}. 
\textcircled{2} \bleudc, \bleucn, and \bleuncs are comparable and always have higher correlation coefficients than other \bleu variants. Among them, the \bleudc performs significantly better, %best,
which indicates that sentence-level \bleu with \methodfour is more relevant to human perception. 
This is because \methodfour smooths zero values without inflating the precision compared to \methodtwo and \methodthree (top of  \Fig~\ref{fig:diff_smoothing_method}).

\begin{tcolorbox}
\textbf{Summary.} 
The differences among the \bleu variants could affect the validity of the experiment and conclusion. (1) The \bleu measure should be implemented correctly 
and described precisely, including the calculation level (sentence or corpus) and the smoothing method being used. (2) The comparison of models should be under the same \bleu metric. (3) \bleudc, the sentence-level \bleu with \methodfour, is more relevant to human perception.

\end{tcolorbox}

\subsection{The Impact of Different Pre-processing Operations (\textbf{RQ2})}\label{diff_process_method}

\begin{table}[t]
\renewcommand\tabcolsep{3.5pt}
\centering 
\small
\caption{The results of four code \processing operations. 1 and 0 denotes use and non-use of a certain operation, respectively. Stars * mean statistically significant.}
\begin{tabular}{c  cc|cc |cc|cc}
\toprule
 Model &$R_0$&$R_1$ &$S_0$  &$S_1$&$F_0$ &$F_1$ & $L_0$ & $L_1$ \\
\midrule
\codenn  &7.19& 7.18& 7.18& 7.19& 7.18& 7.19& 7.19& 7.18\\ 
\astattgru &5.91& 5.97& 5.63& 6.26& 5.85& 6.03& 5.81& 6.07 \\ 
\rencos  &21.85& 21.55& 20.91& 22.5& 21.79& 21.62& 21.43& 21.98 \\
\NeuralCodeSum   &12.20& 12.08& 11.65& 12.63& 12.04& 12.24& 11.82& 12.45 \\ 
\midrule
Avg. & 11.79& 11.70&11.34 & \textbf{12.15\textsuperscript{*}}& 11.72 &11.77& 11.56&11.92 \\
\bottomrule
\end{tabular}
\label{tab:avg_score_of_four_operation_in_bleucn}
\end{table}

In order to evaluate the 
individual effect of four different code \processing operations and the effect of their combinations, we train and test the four models (\codenn, \astattgru, \rencos, and \NeuralCodeSum) under 16 different code \processing %operation 
combinations. Note that the model \deepcom is not experimented as it does not use source code directly. 
In the following experiments, we have performed calculations on all metrics. Due to space limitation, we present the scores under \bleudc, which correlates more with human perception. 
All findings \revise{in the following sections} still hold for other metrics, and the omitted results can be found in the online Appendix. 

As shown in \Tab~\ref{tab:avg_score_of_four_operation_in_bleucn}, for all models, performing $S$ (identifier splitting) is always better than not performing it, while it is unclear whether to perform the other three operations. Then, we conduct the two-sided \emph{t-test}~\cite{dowdy2011statistics} and \emph{Wilcoxon-Mann-Whitney test}~\cite{MannWhitney47} to statistically evaluate the difference between using or dropping each operation. 
The significance signs (*) labelled in \Tab~\ref{tab:avg_score_of_four_operation_in_bleucn} mean that the p-values of the statistical tests at 95\% confidence level are less than 0.05. The results confirm that the improvement achieved by performing $S$ is statistically significant, while performing the other three operations does not lead to statistically different results\footnote{The detailed statistical test scores can be found in the online Appendix Tables 11 to 19. }. 
As pointed out in \cite{KarampatsisBRSJ20}, the OOV (out of vocabulary) ratio is reduced after splitting compound words, and using subtokens allows a model to suggest neologisms, which are unseen in the training data. Many studies \cite{AllamanisPS16,GraveJU17,MerityX0S17,bazzi2002modelling,LuongSM13} have shown that the performance of neural language models can be improved after handling the OOV problem. Similarly, the performance of code summarization is also improved after performing $S$.

\begin{table*}[!ht]
\caption{Performance of different code \processing combinations. Bottom 5 in underline, top 5 in bold, and ensemble models in bold and with gray background.}

\renewcommand\tabcolsep{2.5pt}
\centering
\small
\begin{tabular}{c cccccccccccccccc |c}
\toprule
Model & \preoooo & \preoool & \preoolo & \preooll & \preoloo & \preolol & \preollo & \preolll & \prelooo & \prelool & \prelolo & \preloll & \prelloo & \prellol & \prelllo & \prellll &Ensemble \\
\cmidrule(lr){1-17}\cmidrule(lr){18-18}
    \codenn &\underline{7.06} $\left(6.37\%  \downarrow \right)$ & 7.10& \underline{6.98}& \textbf{7.25}& \textbf{7.54}& \underline{7.01}& \textbf{7.43}& 7.06& 7.22& 7.19& 7.24& \textbf{7.40}& 7.06& \textbf{7.34}$\left(5.16\%  \uparrow \right)$& \underline{7.02}& \underline{7.05} &\cellcolor{lightgray!40}\textbf{10.64}\\ 
    
    \astattgru &\underline{5.67}$\left(14.99\%  \downarrow \right)$ & \underline{5.65}& \underline{5.44}& \underline{5.48}& \textbf{6.17}& \textbf{6.67}& \textbf{6.28}& \textbf{6.41}& 5.84& 5.83& \underline{5.30}& 5.81& 5.79& \textbf{6.62}$\left(24.91\% \uparrow \right)$& 6.03& 6.09 &\cellcolor{lightgray!40}\textbf{11.28}\\ 
    
    \rencos &\underline{20.21}$\left(16.52 \% \downarrow \right)$& \underline{20.35}& 21.28& \underline{21.01}& 21.52& \textbf{23.37}& \textbf{22.25}& \textbf{22.45}& \underline{20.91}& \underline{20.96}& 21.20& 21.33& 21.42& \textbf{24.21}$\left(19.79\%  \uparrow \right)$& \textbf{22.62}& 22.15 &\cellcolor{lightgray!40}\textbf{24.21}  \\ 
    
    \NeuralCodeSum &\underline{11.22}$\left(17.92\%  \downarrow \right)$& 11.95& \underline{11.12}& 12.07& 12.06& \textbf{13.30}& 12.12& \textbf{12.82}& 11.87& \underline{11.51}& \underline{11.78}& \underline{11.64}& \textbf{12.34}& \textbf{13.67}$\left(22.93 \% \uparrow \right)$& 12.09& \textbf{12.67} & \cellcolor{lightgray!40}\textbf{19.90} \\ 
\bottomrule
\end{tabular}
\label{tab:diff_code_pre_dc}
\end{table*}

Next, we evaluate the effect of different combinations of  operations and show the result in \Tab~\ref{tab:diff_code_pre_dc}. For each model, we mark the bottom 5 in underline, the top 5 in bold. We can find that:
\begin{itemize}
        \item Different \processing operations can affect the overall performance by a noticeable margin.
    \item \prellol is a recommended code \processing method, as it is top 5 for all approaches. \preoooo is the not-recommended code \processing method, as it is bottom 5 for all approaches.
    \item The \rank of performance for different models are generally consistent under different code \processing settings.
\end{itemize}

\textbf{An exploration experiment}  From \Tab~\ref{tab:diff_code_pre_dc}, we can see that there is no dominated \processing combination across these approaches. We conduct a simple exploratory experiment that aggregates four different \processing: \prellol, \preolol, \preollo, and \preolll, which mostly perform better than other combinations on the four approaches. 
We use the stacking-based technique~\cite{abs-2107-11423} (the online Appendix \Fig 1) to aggregate the component models. In detail, ensemble components have the same network structure but the input data is processed by different \processing combinations. The result is shown in the last column of \Tab\ref{tab:diff_code_pre_dc}. 
We can see that in general, the ensemble model performs better than the single models, indicating that different \processing combinations may contain complementary information that can improve the final output through ensemble learning.
\begin{tcolorbox}
\textbf{Summary.} 
Code \processing has a large impact on performance (-18\% to +25\%). And, there is no dominated \processing combination for different approaches.
In addition, a simple ensemble model on the different \processing can boost the performance of the model. We share the implementations of 4 code \processing operations and 16 combinations for the convenience of follow-up research.
\end{tcolorbox}

\subsection{How Do Different Characteristics of Datasets Affect the Performance?(\textbf{RQ3})} 
\label{diff_dataset}
\begin{table}[t]
\renewcommand\tabcolsep{4pt}
\centering 
\small
\caption{Performance in different datasets. \revise{Statistically significant ($p \ll 0.05$) results are marked with star *.}}
    \begin{tabular}{c cccc}
  \toprule
  \multirow{2}{*}{Model} & \multicolumn{3}{c}{Dataset}\\
  \cmidrule(r){2-4}
      &\tlcori  & \fcmfiltered &\csnfiltered \\
    \midrule
        \codenn &28.24$\pm$0.19 &12.64$\pm$0.13 &3.32$\pm$0.09 \\ 
         \deepcom &15.65$\pm$2.12 &9.12$\pm$0.03 &1.98$\pm$0.30\\ 
        \astattgru &25.90$\pm$0.79 &15.58$\pm$0.11 &5.01$\pm$0.27 \\ 
        \rencos&\textbf{42.46$\pm$0.05*} &15.47$\pm$0.00 &6.65$\pm$0.05\\
        \NeuralCodeSum&39.50$\pm$0.23 &\textbf{18.07$\pm$0.46*} &\textbf{6.66$\pm$0.51}\\
        \midrule
        \avg &30.35$\pm$9.70 &14.17$\pm$3.05 &4.72$\pm$1.85  \\
    \bottomrule
    \end{tabular}

\label{tab:diff_datasets_dc}
\end{table}

To answer RQ3, we evaluate the five approaches on the three base datasets:  \tlcori, \csnfiltered, and \fcmfiltered. 
From \Tab~\ref{tab:diff_datasets_dc}, we can find that:
\begin{itemize}
    \item The performance of the same model is different on different datasets.
    \item The \rank among the approaches does not preserve when evaluating them on different datasets. For instance, \rencos outperforms other approaches in \tlcori but is worse than \astattgru and \NeuralCodeSum in \fcmfiltered. \codenn performs better than \astattgru on \tlcori, but \astattgru outperforms \codenn in the other two datasets.
    \item The average performance of all models on \tlcori is better than the other two datasets, although \tlcori is much smaller (about 96\% less than \fcmfiltered and 84\% less than \csnfiltered).
    \item The average performance of \fcmfiltered is better than that of  \csnfiltered.
\end{itemize}

\begin{tcolorbox}
\textbf{Summary.} To more comprehensively evaluate different models, it is recommended to use multiple datasets, as the \rank among models can be inconsistent on different datasets.
\end{tcolorbox}

Since there are many factors that make the three datasets different, 
in order to further explore the reasons for the above 
results in-depth, we use the controlled variable method to study from three aspects:  corpus sizes, \datasplit ways, and duplication ratios.

\subsubsection{\underline{The impact of different corpus sizes}}

We evaluate all models on two groups (one group contains \csnmtdmed and \csnmtdsmall, the other group contains \fcmmtdlarge, \fcmmtdmed and \fcmmtdsmall). Within each group, the test sets are the same, the only difference is in the corpus size. 

The results are shown in \Tab~\ref{tab:diff_size}. We can find that the \rank between models can be generally preserved on different corpus sizes.
Also, as the size of the training set becomes larger, the performance of the five approaches improves in both groups, which is consistent with the findings of previous work~\cite{AlonBLY19}. 
We can also find that, compared to other models, the performance of \deepcom does not improve significantly when the size of the training set increases. We suspect that this is due to the high OOV ratio, which affects the scalability of the \deepcom model ~\cite{HellendoornD17,KarampatsisBRSJ20}, as shown in the bottom of \Tab~\ref{tab:diff_size}. 
\deepcom uses only SBT and represents an AST node as a concatenation of the type and value of the AST node, resulting in a sparse vocabulary. Therefore, even if the training set becomes larger, the OOV ratio is still high. Therefore, \deepcom could not fully leverage the larger datasets.

\begin{tcolorbox}
\textbf{Summary.} If additional data is available, one can enhance the performance of models by training with more data since the performance improves as the size of the training set becomes larger. The ranking among models can be generally preserved on different corpus sizes.
\end{tcolorbox}

\begin{table*}[t]
\centering 
\small
\caption{The results of different corpus sizes. \revise{Statistically significant ($p \ll 0.05$) results are marked with star *.}}
    \begin{tabular}{cccc cc }
  \toprule
   Model & \fcmmtdsmall &\fcmmtdmed &\fcmmtdlarge & \csnmtdsmall &\csnmtdmed\\
     \cmidrule(lr){1-1} \cmidrule(lr){2-4}\cmidrule(lr){5-6} 
        \codenn &10.37$\pm$0.17 &14.76$\pm$0.17 &18.68$\pm$0.26 &5.20$\pm$0.01 &12.71$\pm$0.23\\ 
        \deepcom&8.99$\pm$0.06 &10.87$\pm$0.20 &11.65$\pm$0.36 &7.57$\pm$0.74 &7.85$\pm$1.07\\ 
        \astattgru &12.86$\pm$0.64 &18.15$\pm$0.05 &21.73$\pm$0.11 &5.89$\pm$0.12 &15.83$\pm$0.17\\ 
        \rencos&14.24$\pm$0.12 &21.97$\pm$0.08 &23.81$\pm$0.04 &7.36$\pm$0.08 &19.56$\pm$0.03\\
        \NeuralCodeSum &\textbf{14.70$\pm$0.19} &\textbf{23.10$\pm$0.32*} &\textbf{29.03$\pm$0.32*} &\textbf{9.07$\pm$0.20*} &\textbf{25.17$\pm$0.39*}\\
\midrule
     OOV Ratio of \deepcom &91.90\% &88.94\% &88.32\% &91.49\% & 85.81\%\\
     OOV Ratio of Others &63.36\% &53.09\% &48.60\% &60.99\%  &34.00\% \\
    \bottomrule

    \end{tabular}
\label{tab:diff_size}
\end{table*}
% \vspace{-20pt}
\subsubsection{\underline{The impact of different data splitting methods}}

In this experiment, we evaluate the five approaches on two groups (one group contains \fcmpjtlarge and \fcmmtdlarge and another contains \csnpjtmed, \csnclsmed, \csnmtdmed). Each group only differs in \datasplit ways. 
From \Tab~\ref{tab:diff_split_level},  
we can observe that all approaches perform differently in different \datasplit ways, and they all perform better on the dataset split by method than by project. This is because similar tokens and code patterns are used in the methods from the same project~\cite{PanthaplackelNG20,leclair2019recommendations,abs-1912-02972}. 
In addition, when the data splitting ways are different, the rankings between various approaches remain basically unchanged, which indicates that it would not impact comparison fairness across different approaches whether or not to consider multiple data splitting ways.

\begin{tcolorbox}
\textbf{Summary.} Different data splitting methods can significantly affect the 
performance of all models. However, the \rank of the model remains basically unchanged. Therefore, if data availability or time is limited, it is also reliable to evaluate the performance of different models under only one data splitting method. 
\end{tcolorbox}

\subsubsection{\underline{The impact of different duplication ratios}}
\begin{figure}[t]
    \centering
    \includegraphics[width=1\linewidth]{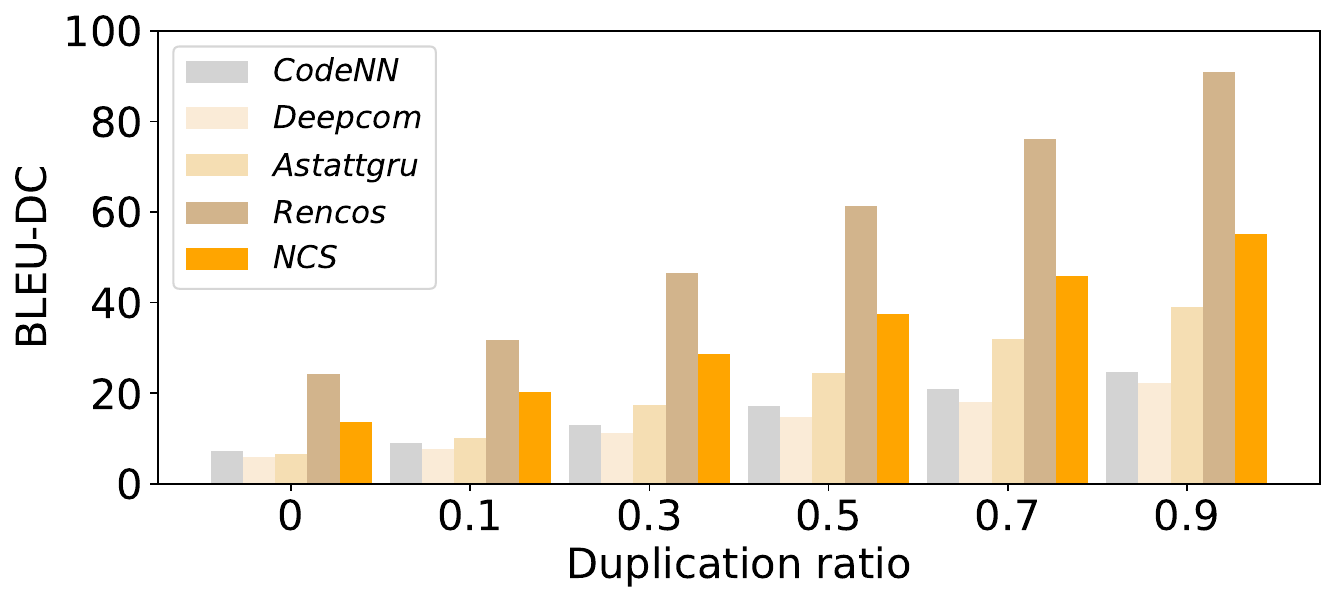}
    \caption{The results of different duplication ratios.}
    \label{fig:diff_dup_ratio}
\end{figure}

To simulate scenarios with different code duplication ratios, we construct synthetic test sets from \tlcdedup by adding random samples from the training set to the test set. Then, we train the five models using the same training set and test them using the synthetic test sets with different duplication ratios (i.e., the test sets with random samples).
From the results shown in \Fig~\ref{fig:diff_dup_ratio},
we can find that:
\begin{itemize}
    \item The \bleu scores of all approaches increase as the duplication ratio increases.
    
    \item The score of the model \rencos increases significantly when the duplication ratio increases. We speculate that the reason should be the duplicated samples being retrieved back by the retrieval module in \rencos. Therefore, retrieval-based models could benefit more from code duplication.
    
    \item In addition, the \rank of the models is not preserved with different duplication ratios. For instance, \codenn outperforms \astattgru without duplication and is no better than \astattgru on other duplication ratios. 
\end{itemize}

\begin{tcolorbox}
\textbf{Summary.} To evaluate the performance of neural code summarization models, it is recommended to use deduplicated datasets so that the generalization ability of the model %itself
can be tested. However, in real scenarios, duplications are natural. Therefore, we suggest evaluating models under different duplication ratios. Moreover, it is recommended to consider incorporating retrieval techniques to improve the performance especially when code duplications exist.
\end{tcolorbox}

\begin{table*}[t]
\renewcommand\tabcolsep{4pt}
\centering \small
\caption{The results in different \datasplit methods. \revise{Statistically significant ($p \ll 0.05$) results are marked with star *.}}
\begin{tabular}{p{3.1cm} ccccc}
  \toprule
\centering {Model} &\csnpjtmed   & \csnclsmed  &\csnmtdmed &\fcmpjtlarge &\fcmmtdlarge\\
    % \midrule
     \cmidrule(lr){1-1} \cmidrule(lr){2-4}\cmidrule(lr){5-6} 
        \centering {\codenn} &3.32$\pm$0.09 &9.57$\pm$0.15 &12.71$\pm$0.23 &12.64$\pm$0.13 &18.68$\pm$0.26\\ 
        \centering {\deepcom} &1.98$\pm$0.30 &6.14$\pm$0.12 &7.85$\pm$1.07 &9.12$\pm$0.03 &11.65$\pm$0.36 \\
        \centering{\astattgru}  &6.86$\pm$3.07 &11.72$\pm$0.41 &15.83$\pm$0.17 &15.58$\pm$0.11 &21.73$\pm$0.11\\
        \centering {\rencos} &6.65$\pm$0.05 &14.37$\pm$0.03 &19.56$\pm$0.03 &15.47$\pm$0.00 &23.81$\pm$0.04\\
       \centering { \NeuralCodeSum} &\textbf{ 6.66$\pm$0.51} &\textbf{17.96$\pm$0.23*} &\textbf{25.17$\pm$0.39*} &\textbf{18.07$\pm$0.46*} &\textbf{29.03$\pm$0.32*}\\
    \midrule
   \centering{OOV Ratio}  &48.74\% &35.38\% & 34.00\%  &57.56\%  &48.60\% \\
    \bottomrule
    \end{tabular}
\label{tab:diff_split_level}
\end{table*}

 We observe that even when we control all three factors (splitting methods, duplication ratios, and dataset sizes),  the performance of the same model still varies greatly between different datasets\footnote{The results are given in the online Appendix Tables 61 to 69 due to space limitation.}. This indicates that the differences in training data may also be a factor that affects the performance of code summarization. We leave it to future work to study the impact of data differences.

\section{Threats to Validity}
We have identified the following main threats to validity:

\textbf{\emph{Programming languages.}} We only conduct experiments on Java datasets. Although in principle, the models and experiments are not specifically designed for Java, more evaluations are needed when generalizing our findings to other languages. In the future, we will extend our study to other programming languages.

\textbf{\emph{The quality of summaries.}}  
The summaries in all datasets are collected by extracting the first sentences of Javadoc. Although this is a common practice to place a method's summary at the first sentence according to the Javadoc guidelines\footnote{\url{http://www.oracle.com/technetwork/articles/java/index-137868.html}}, there might still be 
some incomplete or mismatched summaries in the datasets.

\textbf{\emph{Models evaluated.}} We covered all representative models with different characteristics, such as Transformer-based and RNN-based models, single-channel and multi-channel models, models with and without retrieval techniques. However, other models that we are out of our study may still cause our findings to be untenable.

\textbf{\emph{Human evaluation.}} 
We use two different ways (arithmetic and geometric average) to aggregate the sentence-level human scores as a corpus-level human score. 
The aggregation method may threaten our conclusion. We will explore other ways to assess corpus-level quality in human evaluation.

\section{Related Work}
Code summarization plays an important role in comprehension, reusing and maintenance of program.
 Some surveys~\cite{nazar2016summarizing,song2019survey,zhu2019automatic} provided a taxonomy of code summarization methods and discussed the advantages, limitations, and challenges of existing models from a high-level perspective. Especially, Song et al.~\cite{song2019survey} also provided a discussion of the evaluation techniques being used in existing methods.  
Gros et al.~\cite{GrosSDY20} described an analysis of several machine learning approaches originally designed for the task of natural language translation for the code summarization task. They also observed that different datasets were used in existing work and different metrics were used to evaluate different approaches.
Allamanis et al.~ \cite{Allamanis19} explored the effect of code duplication and concluded that the performance of the technique is sometimes overestimated when evaluated on the duplicated dataset.
LeClair et al.~ \cite{leclair2019recommendations} conducted the experiment of a standard NMT algorithm from two aspects: splitting strategies (splitting the dataset by project or by method) and a clean approach, and proposed the guidelines
for building datasets based on experiment results. 
Some studies~\cite{StapletonGLEWL020,RoyFA21} conducted a human study and concluded that \bleu is not correlated to human quality assessments when measuring one generated summary. Roy et al.~ \cite{RoyFA21} also re-assessed and interpreted other automatic metrics for code summarization. 
Our work differs from previous work in that we not only observe the inconsistent usage of different BLEU metrics but also conduct dozens of experiments on the five models and explicitly confirm that the inconsistent usage can cause severe problems in evaluating/comparing models. Besides, we perform a human evaluation to provide additional findings, e.g., which \bleu metrics correlate with human perception the most. 
Moreover, we explore factors affecting model evaluation, which have not been systematically studied before, such as dataset size, dataset split methods, code pre-processing operations, etc. Different from the surveys, we provide extensive experiments on various datasets for various findings and corresponding discussions. Finally, we consolidate all findings and propose actionable guidelines for evaluating code summarization models.

\section{Conclusion}
In this paper, we conduct an in-depth analysis of recent neural code summarization models. We have investigated several aspects of model evaluation: evaluation metrics, code \processing operations, and datasets. Our results point out that all these aspects have  large impact on evaluation results. Without a carefully and systematically designed experiment, neural code summarization models cannot be fairly evaluated and compared.
Our work also suggests some actionable guidelines including:
(1) Reporting BLEU metrics explicitly (including sentence or corpus level, smoothing method, \nltk version, etc). \bleudc, which correlates more with human perception, can be selected as the evaluation metric. 
(2) Using proper (and maybe multiple) code \processing operations.
 (3) Considering the dataset characteristics when evaluating and choosing the best model. 
 We build a shared code summarization toolbox
 containing the implementation of \bleu variants, code \processing operations, datasets, the implementation of baselines, and all experimental results.
 We believe the results and findings we obtained can be of great help for practitioners and researchers working on this interesting area. 

For future work, we will extend our study to programming languages other than Java. We will design an automatic evaluation metric which is more correlated to human perception. We will also explore more attributes of datasets. 
Furthermore, we plan to extend the study 
to other text generation tasks in software engineering such as commit message generation.

To facilitate reproducibility, our code and data are available at \url{https://github.com/DeepSoftwareAnalytics/CodeSumEvaluation}. 

\section{Acknowledgement}
We thank reviewers for their valuable comments on this work. This research was supported by \revise{National Key R\&D Program of China (No.2017YFA0700800)}.
We would like to thank Jiaqi Guo for his valuable suggestions and feedback. We also thank the participants of our human evaluation for their time. 

\balance
\bibliographystyle{ACM-Reference-Format}
\bibliography{ref}

\end{document}